\documentclass[preprint]{revtex4-1}
\usepackage{graphicx}
\usepackage{float}
\usepackage{xcolor}
\usepackage{soul}

\begin{document}


\title{Revealing and exploiting hierarchical material structure through complex atomic networks}

\author{Sebastian E. Ahnert}
\affiliation{Theory of Condensed Matter Group, Cavendish Laboratory, University of Cambridge, JJ Thomson Avenue, Cambridge CB3~0HE, United Kingdom}
\affiliation{Sainsbury Laboratory, University of Cambridge, Bateman Street, Cambridge CB2~1LR, United Kingdom}

\author{William P. Grant}
\affiliation{Theory of Condensed Matter Group, Cavendish Laboratory, University of Cambridge, JJ Thomson Avenue, Cambridge CB3~0HE, United Kingdom}

\author{Chris J.\ Pickard\\ (\url{cjp20@cam.ac.uk})}
\affiliation{Department of Materials Science \& Metallurgy, University of Cambridge, 27 Charles Babbage Road, Cambridge CB3~0FS, United Kingdom}
\affiliation{Advanced Institute for Materials Research, Tohoku University 2-1-1 Katahira, Aoba, Sendai, 980-8577, Japan}


\date{\today}

\begin{abstract}
One of the great challenges of modern science is to faithfully model, and understand, matter at a wide range of scales. Starting with atoms, the vastness of the space of possible configurations poses a formidable challenge to any simulation of complex atomic and molecular systems. We introduce a computational method to reduce the complexity of atomic configuration space by systematically recognising hierarchical levels of atomic structure, and identifying the individual components. Given a list of atomic coordinates, a network is generated based on the distances between the atoms. Using the technique of modularity optimisation, the network is decomposed into modules. This procedure can be performed at different resolution levels, leading to a decomposition of the system at different scales, from which hierarchical structure can be identified. By considering the amount of information required to represent a given modular decomposition we can furthermore find the most succinct descriptions of a given atomic ensemble. Our straightforward, automatic and general approach is applied to complex crystal structures. We show that modular decomposition of these structures considerably simplifies configuration space, which in turn can be used in discovery of novel crystal structures, and opens up a pathway towards accelerated molecular dynamics of complex atomic ensembles. The power of this approach is demonstrated by the identification of a possible allotrope of boron containing 56 atoms in the primitive unit cell, which we uncover using an accelerated structure search, based on a modular decomposition of a known dense phase of boron, $\gamma$-B$_{28}$.
\end{abstract}


\maketitle

\section*{Introduction}

Considerable growth in computational power and its ubiquity has been coupled with the development of efficient algorithms \cite{car1985unified,payne1992iterative} and their implementation in robust \cite{clark2005first,giannozzi2009quantum,kresse1996efficient} and reliable \cite{lejaeghere2016reproducibility} computer codes. This has permitted the first principles, quantum mechanical (through density functional theory - DFT \cite{hohenberg1964inhomogeneous,kohn1965self,parr1995density}), treatment of material \cite{hasnip2014density}, chemical \cite{zhao2008density}, and biological systems \cite{cole2010protein} of increasing complexity. High throughput computations can be performed directly on pre-existing data of more modest complexity, or some modification of it \cite{jain2016computational}, or as a way to sample configuration space to discover previously unknown structures \cite{pickard2006high,Pickard:2011bu,oganov2006crystal,wang2012calypso}. Together, these approaches offer a route to computational materials discovery \cite{needs2016perspective}.

In parallel, increasing computational power and an abundance of data has given rise to another rapidly expanding field - the study of complex networks \cite{Watts:1998vz,Barabasi:1999uu,Albert:2002wu,Newman:2010nw}. A key reason for its success is the fact that related mathematical approaches can be applied to a wide range of real-world network data across many academic disciplines. The structure of networks can be studied at a variety of resolutions. Local measures of connectivity can quantify the topological properties of individual nodes or edges. Global measures, calculated across the entire network, such as the average shortest path length \cite{Watts:1998vz}, can help us to compare networks as a whole. Between these two extremes however lies an entire field of research that searches for meaningful descriptions of intermediate structures, such as ``cliques'' \cite{Palla:2005cj}, ``communities'' \cite{Newman:2004ep}, and ``rich clubs'' \cite{Colizza:2006kz}, among others. These are sets of nodes or edges which are particularly densely connected, or which share some other defining topological feature. Many definitions of such structures have been put forward \cite{Newman:2004ep,Palla:2005cj,Colizza:2006kz,Blondel:2008do,Arenas:2008hq,Ahn:2010dj}. Here we select an approach that is particularly good at detecting hierarchical modularity \cite{Arenas:2008hq} and apply it to atomic networks, which until now have received scant attention from the networks research community, beyond the study of proteins \cite{Delvenne:2010iw,Delmotte:2011bn,Yaliraki:2014wj}. Our aim is to provide an automated coarse-graining of the atomic structures of crystal structures. The simplification of the space of possible configurations of complex atomic systems has the potential to vastly accelerate the process of computational materials discovery, among other tasks that can benefit from automatic coarse graining based on hierarchical decomposition. We illustrate this through the identification of a possible new allotrope of boron.

\section*{Results}

\subsection*{Determining the modularity of atomic networks}

For atomic structures of a single atomic species we can generate an unweighted network of atoms by simply imposing a threshold $d^*$ on the interatomic distance and connecting atoms that are closer to each other than this threshold distance. The communities in this network can then be extracted by using the algorithm of Arenas et al. \cite{Arenas:2008hq}, as discussed below.

\begin{figure}[]
\centering
\includegraphics[width=0.5\textwidth]{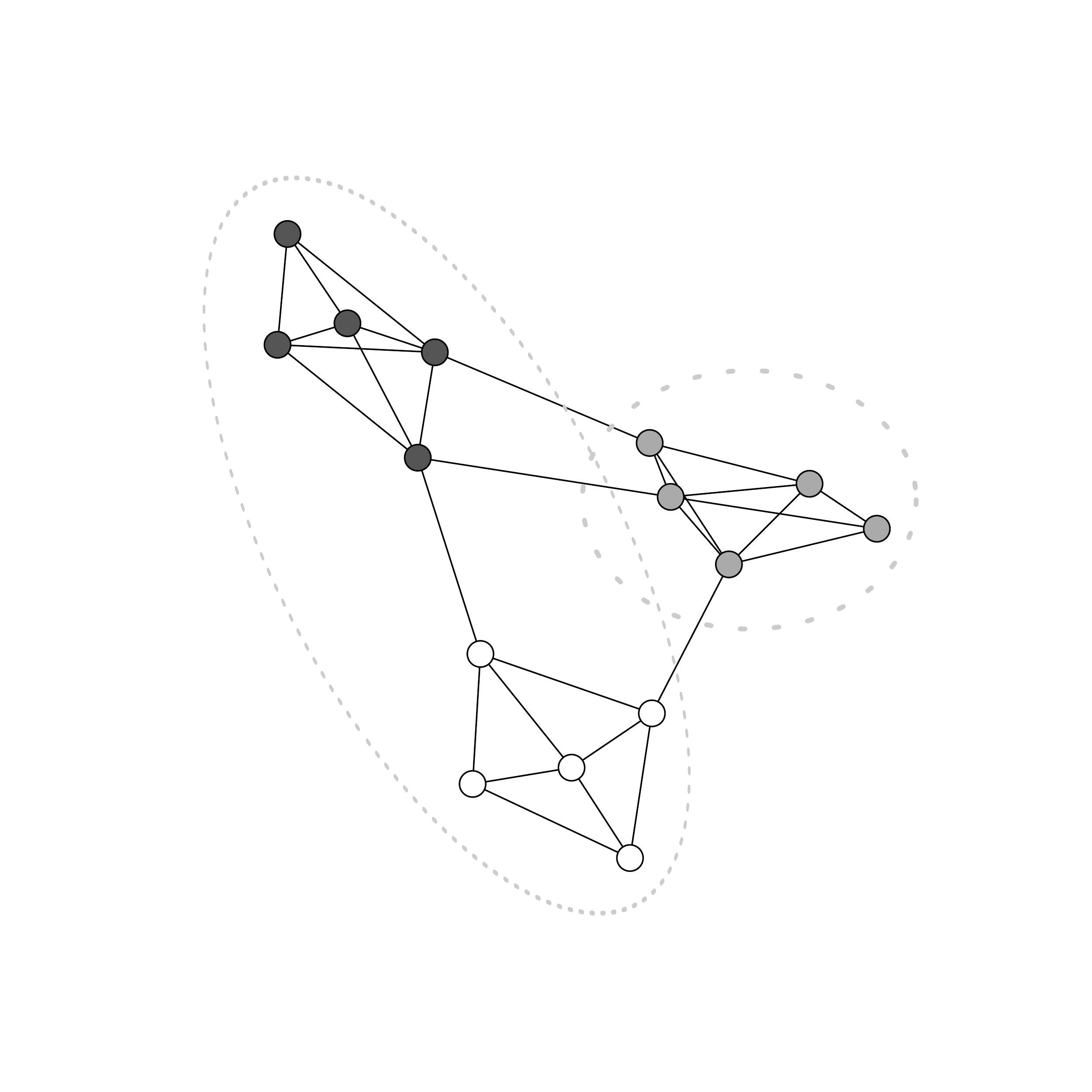}
\caption{A network with three communities of 5 nodes each. Let the white, grey and black communities be $C_1$, $C_2$, $C_3$. Then $A_{ss}$, the number of internal edges, is $\{8, 9, 9\}$. The sum of degrees of each community, $A_s$ is $\{18, 21, 21\}$, and the total number of edges is 30, giving a final value for Q of 0.532. An alternative partition, indicated by the dotted lines, has $A_{ss}= \{18, 9\}$ and $A_{s} = \{39, 21\}$, giving $Q=0.355$. As expected, the less intuitive partitioning gives a lower Q value.}\label{examplenetwork}
\end{figure}

The extent to which a network has well-defined community structure can be quantified with a metric known as the modularity \cite{newman04} (Fig. \ref{examplenetwork}). This is defined as the fraction of edges that run between nodes of the same community, minus the expected fraction if the edges of the network were positioned randomly:

\begin{equation}
Q = \frac{1}{2A} \sum_{i,j} (A_{ij}-P_{ij}) \; \delta(C_i,C_j)
\end{equation}
\noindent
Here $i,j \in [1,n] $, where $n$ is the number of nodes. $\delta(C_i,C_j) = 1$ if nodes $i$ and $j$ belong to the same community, and $0$ otherwise, and $A_{ij}$ and $P_{ij}$ are the adjacency matrices of the network and of the null model (the randomised network), respectively. $A = \frac{1}{2}\sum_{i,j} A_{ij}$ gives the total number of edges in the network.

This metric relies on the concept that a random graph is not expected to exhibit community structure. As such, the quality of the proposed community structure can be quantified by the difference between the network and the null model. The degree of a node $i$ is defined as the number of edges of the node, $A_i = \sum_{j} A_{ij} \; $. Choosing the null model to have the same degree distribution as the target network gives:

\begin{equation}
Q = \frac{1}{2A} \sum_{i,j} (A_{ij}- \frac{A_i A_j}{2A}) \; \delta(C_i,C_j)
\end{equation}

\noindent
This can be rewritten as a sum over the $M$ communities of the network:

\begin{equation}
Q = \sum_{s=1}^{M} \left( \frac{A_{ss}}{A} - \left( \frac{A_s}{2A} \right)^2  \right)
\end{equation}

\noindent
Where for unweighted networks $A_{ss}$ is the number of edges within community $s$, and $A_s$ is the sum of the degrees of nodes within community $s$. (Equivalently, $A_s$ is the number of edges exiting the community + $2A_{ss}$.) This metric allows community detection to be recast as an optimisation problem; maximising Q minimizes the number of edges between communities.  However, naive optimisation of the modularity has been shown to have a fundamental resolution limit \cite{fortunato2007}.
Communities smaller than a certain size $A_{ss}^{min}$ will not be detected. This threshold depends on the total number of edges in the network:

\begin{equation}
A_{ss}^{min} = \sqrt{\frac{A}{2}} - 1
\end{equation}

\noindent
This resolution limit arises due to the explicit dependence on the number of edges within the null model. This introduces a preferred size for the communities in the network.

The algorithm of Arenas et. al \cite{Arenas:2008hq} utilises this bound on the size of detectable communities. By adding a self-loop of strength $w$ to each node in network, $\mathbf{A} \rightarrow \mathbf{A} + w\mathbf{I}$, the resolution limit inequality becomes:

\begin{equation}
A_{ss} < \frac{1}{2}\left(\sqrt{2A+Nw} - N_sw -2 \right)
\end{equation}

\noindent
Where $N_s$ is the number of nodes in community $s$, and $N$ the total number of nodes. By varying the effective resolution limit, the scale of the communities extracted can be varied. As $w$ is increased, modularity optimisation will result in an increasingly fragmented decomposition. By optimising modularity for a range of this parameter $w$, and across a range of threshold values $d^*$, one obtains a variety of hierarchical decompositions into modules (see the section ``Relax and Shake Algorithm'' for details of the optimisation algorithm). The simplest quantity one can establish across this two-dimensional space is the number of modules. Regions across which this value is stable represent more meaningful modules that may have real physical or biological meaning as rigid clusters or units of protein architecture. 

Pauling's rule of parsimony suggests that the number of different kind of constituents in a crystal is small \cite{pauling1929principles}.  This suggests that in complex crystal structures, in which modules are likely to exhibit a degree of symmetry, it makes sense to minimise a more sophisticated quantity, namely the information content of a structure. The identification of modules corresponds to a compression if these modules contain symmetries or if the same module appears multiple times. We can calculate the amount of information $I$ required to describe a given module structure in terms of the global degrees of freedom of that structure, and minimise this quantity over the space of $d^*$ and $w$.

In order to calculate $I$, consider $M$ modules of $M'$ distinct types. The number of modules with one atom only is $0 \le M^* \le M$, and the number of modules with two atoms is $0 \le M^{**} \le M$. To position and rotate the $M$ modules relative to one another we need $6M-6$ degrees of freedom in general, with one degree less for every two-atom module, and three degrees less for every one-atom module. Now consider each of the $M'$ distinct modules: If we have $n_i > 2$ inequivalent atoms in module $i$ we need $3 n_i - 6$ degrees of freedom to describe them. If we have $n_i = 2$ inequivalent atoms in module $i$ we need $3 n_i - 5 = 1$ degree of freedom to describe them, which corresponds to the distance between the atoms. If we have $n_i = 1$ inequivalent atom in module $i$ we need 0 degrees of freedom to describe it internally. The global number of degrees of freedom is then:

\begin{equation}
I = 6M-M^{**}-3M^* - 6 + \sum_{i=1}^{M'} 3 n_i - 6 + 3 \delta_{1n_i} + \delta_{2n_i}
\end{equation}

\noindent
Note that the number of inequivalent atoms $n_i$ depends on the number of atoms in module $i$, as well as its point group symmetry.
If all $N$ atoms are in one module, repeated once, we have $M = 1$, $M' = 1$, $M^* = 0$, $M^{**} = 0$, and $n_i = N$. Hence $I = 3N - 6$, as required. If all $N$ atoms are in $N$ modules of one atom we have $M = N$, $M' = 1$, $M^* = N$, $M^{**} = 0$, $n_i = 1$. Hence in this case also $I = 3N - 6$, as required. The information $I$ can be normalised by the maximum possible value of $3N-6$. This normalised value is used in all the heatmap figures in this manuscript.

The decomposition which has the minimum $I$ gives us the most concise description of a structure. This minimisation of the description length is conceptually related to the idea of algorithmic information theory \cite{kolmogorov1965three,chaitin1982algorithmic,ahnert2010self}, as the symmetry operations and inequivalent atomic positions that form part of the compression can be thought of algorithms which allow us to reconstruct the original atomic structure. The length of the shortest such description is a quantitative measure of the structure's complexity. Because of the presence of crystal symmetries, we need to establish the modularity of the atomic network with high accuracy. In addition, the modularity is highly degenerate; there is a greater than exponential number of distinct possible community structures, and many will have modularity values close to that of the global maximum \cite{good2010}. Moreover, these structures may have very different topologies to that of the true partition, resulting in a large change in the compression achieved. We therefore employ an algorithm similar to that of the 'relax and shake' algorithm \cite{Pickard:2011bu} or zero temperature basin hopping \cite{leary2000global}.

\subsection*{Relax and Shake Algorithm}
The relax and shake (RASH) algorithm uses a repeated series of local modularity optimisations (relax) followed by the assignment of a small number of nodes into random communities (shake), in order to escape local maxima. The local optimisation follows existing work on community detection \cite{blondel2008, massen2005}. The modularity is optimised by moving each node in the network to the community of the neighbouring node resulting in the highest increase in the modularity (if $>0$). This is then repeated until no further local optimisations increase the modularity.
Following the local optimisation, a subset of the nodes (10\%) are shaken into other communities within the network, and the local optimisation repeated. This continues until 200 consecutive relax-and-shake iterations have failed to improve the modularity. As an additional check on the solution, the modularity change resulting in merging any two communities is calculated; if this results in a modularity increase, the merge is performed, and the relax-and-shake iteration process is begun again using the new partition. The above can be considered a single optimisation step; following this, a larger subset of the nodes (20\%) are shaken into either pre-existing, or new communities. The optimisation is then performed again. This shake-and-relax is performed until 200 iterations fail to improve the modularity.The whole process is repeated until three consecutive runs have failed to produce a community structure with a higher modularity. The degree of repetition is parameterisable, and allows us to have confidence in the community structure obtained (at the cost of speed).

If we partition a structure of $N$ atoms into $M$ multi-atom modules, so that typically $M << N$, and assume that the modules correspond to rigid clusters, then we reduce the dimensionality of configuration space from $3N-6$ (atomic positions minus global translation and rotation) to $6M-6$ (as we have to specify a relative translation and rotation for each module) or less (if any of the modules have less than three atoms). We will show that this can be exploited in the first principles prediction of crystal structure.

\subsection*{Application to crystal structures}

\subsubsection*{Boron}

\begin{figure}[]
\centering
\includegraphics[width=0.5\textwidth]{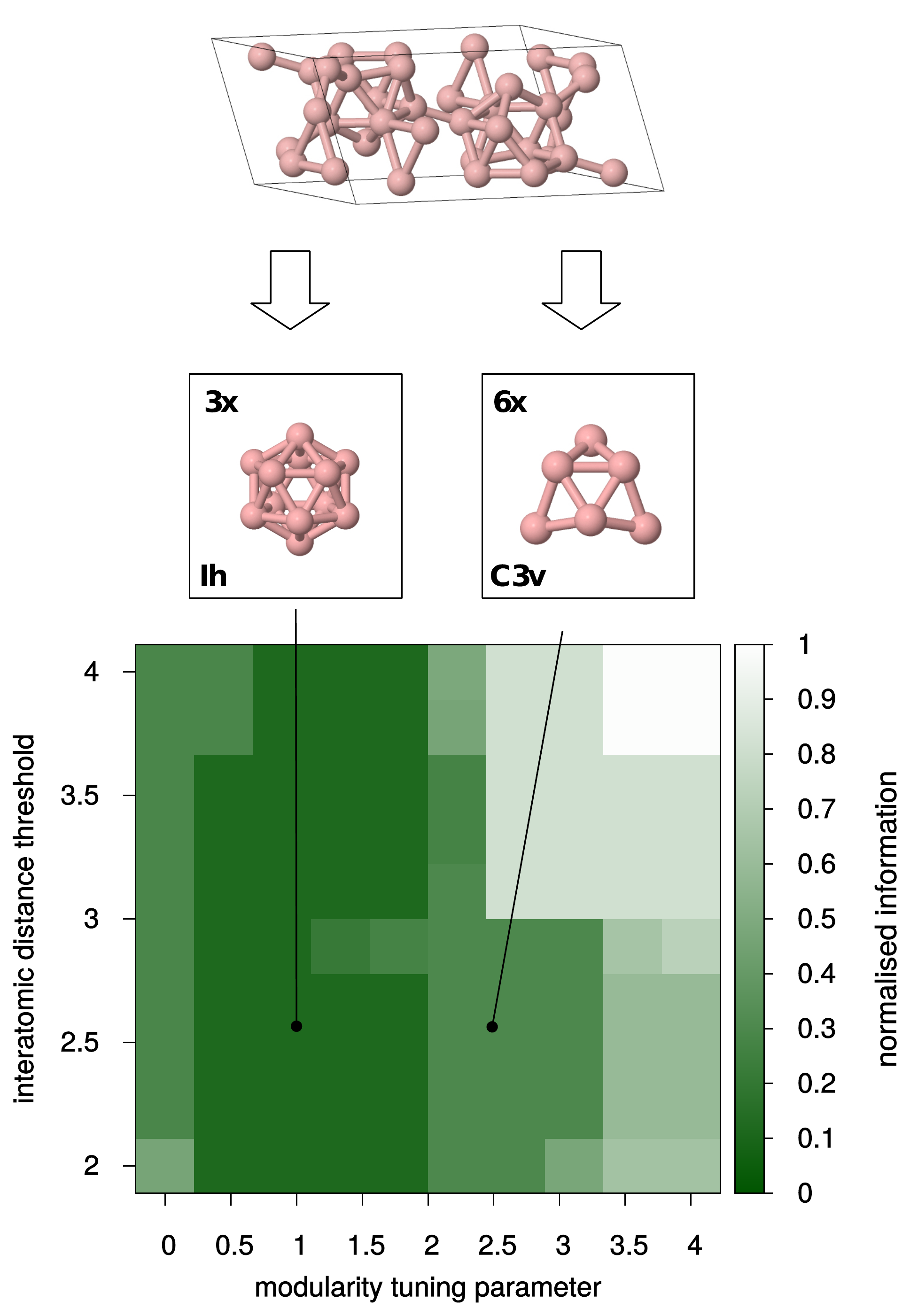}
\caption{The structure of $\alpha$-B$_{12}$ (in the 36 atom hexagonal unit cell) and its modular decompositions. As established by the modularity decomposition the simplest way of describing this structure is as three interconnected icosahedra. For the purpose of comparison we also show an alternative, less efficient decomposition into six units of six atoms with C$_{3v}$ symmetry. These units correspond to halves of the icosahedra. The decomposition that minimises the amount of information $I$ (normalised here by the maximum value of $3N-6$) required to describe the crystal structure, occurs for values around $w_s = 1$ (modularity tuning parameter) and $d^* = 2.5$\AA (interatomic distance threshold). This decomposition occupies a broad plateau (darkest shade of green) in the two-dimensional space of the modularity tuning parameter $w_s$ and the interatomic distance threshold $d^*$.}\label{B12}
\end{figure}

Boron is known to form several allotropes \cite{albert2009boron}, including $\alpha$-B \cite{decker1959crystal}, $\beta$-B \cite{talley1960new,geist1970verfeinerung}, and $\gamma$-B \cite{Wentorf49,oganov2009ionic,zarechnaya2009superhard}. The structure of rhombohedral $\alpha$-B is widely recognised as being made up of interconnected B$_{12}$ icosahedra (see Fig. \ref{B12}). However, because the bonds between different icosahedra are shorter than the bonds within, simple thresholds on bond length will not yield the underlying modular structure of this crystal phase. This observation motivated the development of our current scheme, which, based on network modularity, does yield the scientifically agreed icosahedral modules.

\begin{figure*}[]
\centering
\includegraphics[width=1\textwidth]{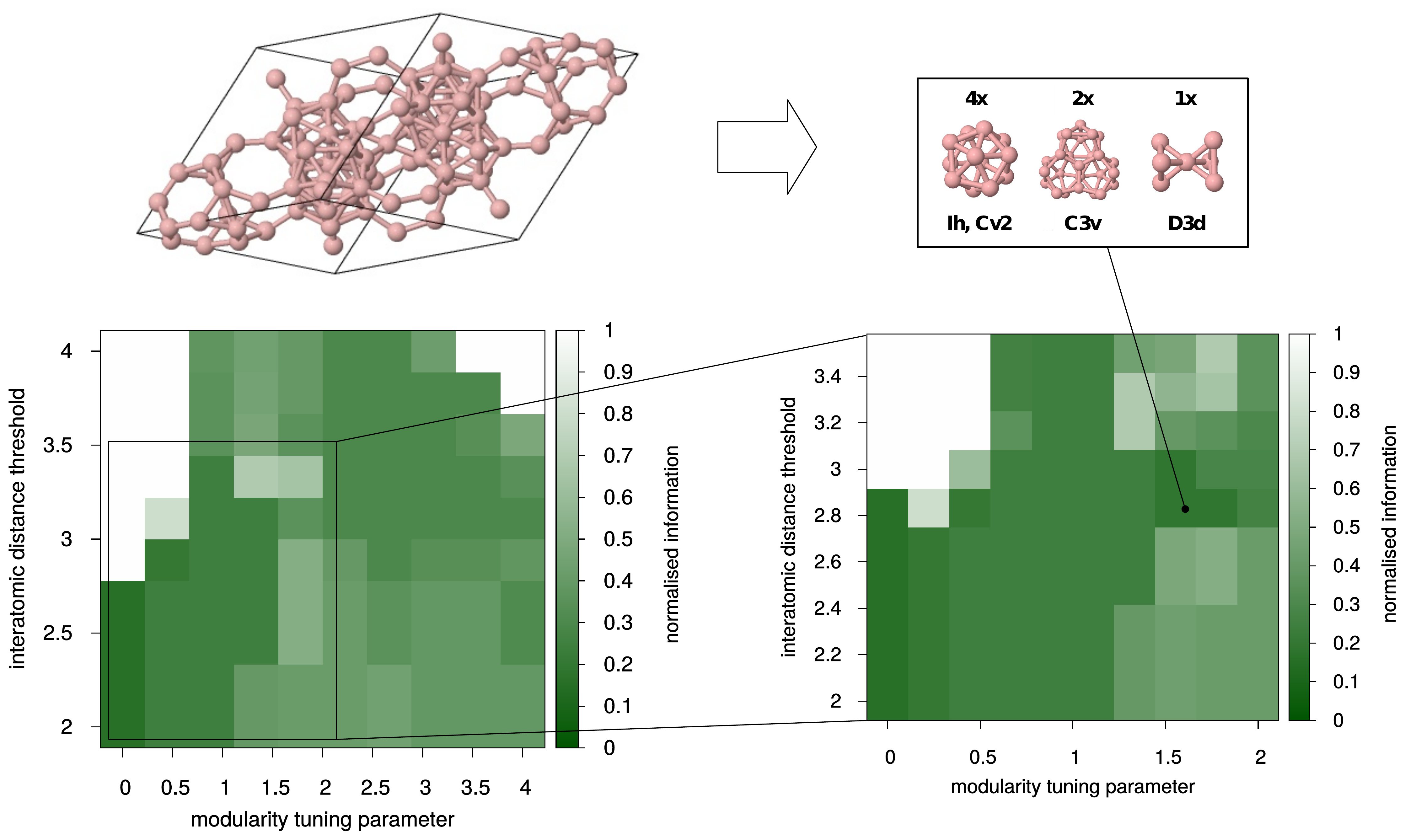}
\caption{The highly complex structure of $\beta$-B$_{105}$ (in the 105 atom rhombohedral unit cell) and its modular decompositions. The complexity of the structure means a more complex landscape of possible decompositions. The module structure that minimises the amount of information $I$ (normalised here by the maximum value of $3N-6$) required to describe the crystal structure, occurs for values around $w_s = 1.66$ and $d^* = 2.82$\AA. This decomposition consists of four icosahedra, two large 25-atom structures with threefold symmetry (C$_{3v}$), and a further module with dihedral symmetry (D$_{3d}$). Two of the four icosahedra have exact icosahedral symmetry, whereas the other two are slightly distorted, and exhibit C$_{2v}$ symmetry.}\label{B105}
\end{figure*}

The structure of $\beta$-B is much more complicated, and various models have been proposed, typically with 105 or 106 atoms per unit cell. We choose the 105-atom model of Geist \emph{et al.}\cite{geist1970verfeinerung}  for further investigation (see Fig. \ref{B105}). Our modularity detection scheme identifies four icosahedra, two larger 25-atom modules with threefold cyclic point group symmetry C$_{3v}$, and one module with threefold dihedral symmetry D$_{3d}$ of seven atoms. Two of the four icosahedra are slightly distorted, resulting in C$_{2v}$ symmetry, rather than icosahedral $I_h$ symmetry. Of course, the decomposition of complex boron structures into compact and symmetric sub units is not unprecedented, see for example Figure 2 in Albert and Hillebrecht \cite{albert2009boron}. We emphasise, however, that our scheme performs the decomposition automatically and is suitable for integration into complex computational workflows.

\begin{figure}[]
\centering
\includegraphics[width=0.5\textwidth]{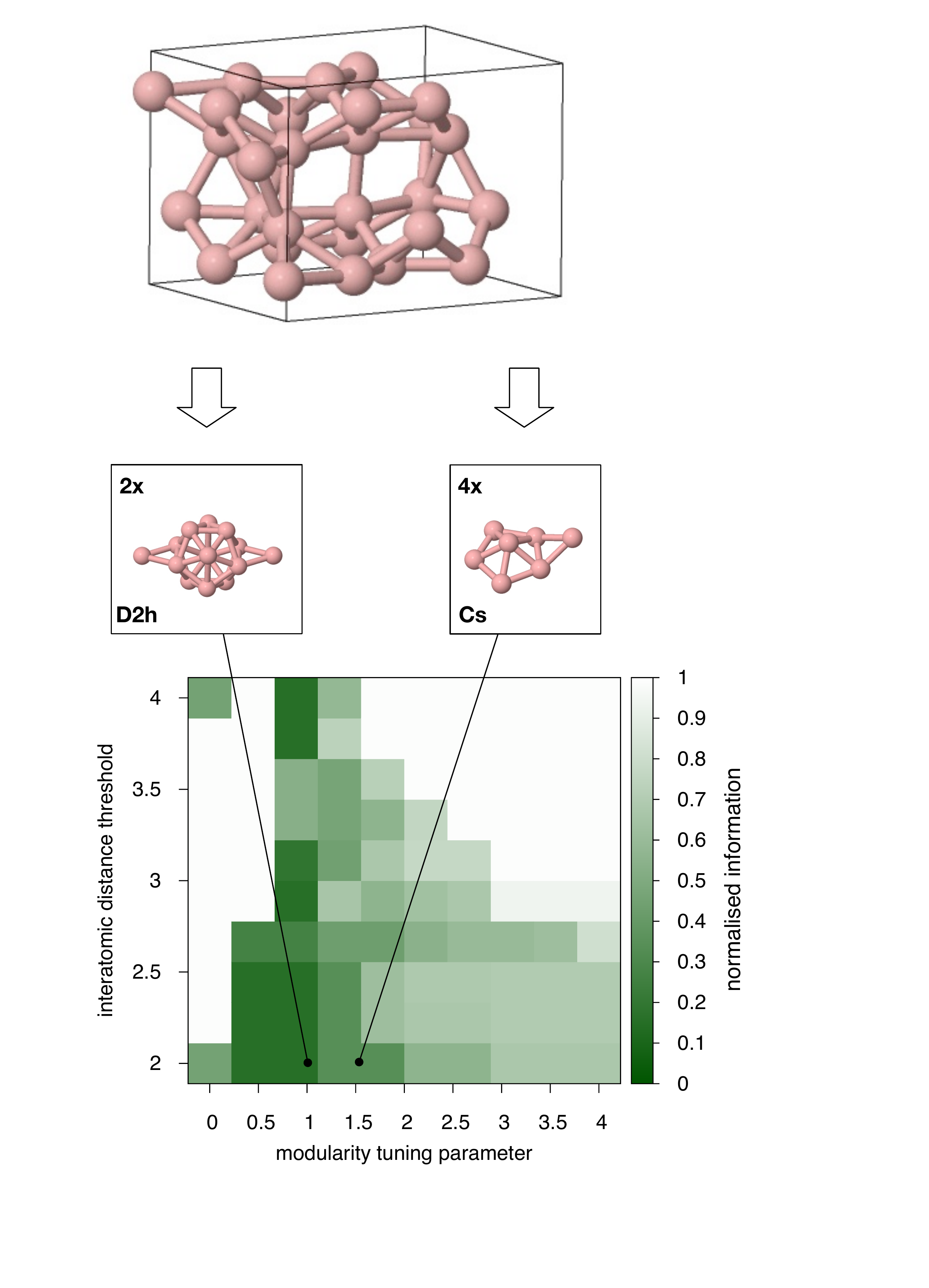}
\caption{The structure of $\gamma$-B$_{28}$ (Pnnm unit cell) and its modular decompositions. The most parsimonious decomposition, which minimises the amount of information $I$  (normalised here by the maximum value of $3N-6$) required to describe the crystal structure, occurs for values around $w_s = 1.0$ and $d^* = 2.0$\AA. This decomposition consists of two identical modules of fourteen atoms with dihedral symmetry (D$_{2h}$), which in turn are each composed of an icosahedron with two adjoining atoms, one on either side. For the purposes of comparison we also show a less efficient decomposition, with C$_s$ symmetry.}\label{B28}
\end{figure}

Recently the structure of a high-pressure phase of boron, $\gamma$-B$_{28}$, has been described in the literature \cite{oganov2009ionic,zarechnaya2009superhard,albert2009boron} (see Fig. \ref{B28}). A $w_s$ versus $d^*$ heat-map of $I$ for the unweighted network reveals a global minimum at $w_s=1.0$ and $d^*=2.0$\AA. This corresponds to two 14-atom modules with D$_{2h}$ symmetry, which are icosahedra plus two atoms on either side. (see Fig. \ref{B28}). Note that this contrasts with the conventional decomposition into two icosahedra and two dimers found in the literature \cite{oganov2009ionic}, which is less favoured in our scheme as it corresponds to a higher value of $I$. Our approach offers a meaningful partition of this structure into modules, providing insight into the organisation and visualisation of this structure and opening the door to the systematic exploration of the structure space that neighbours this $\gamma$-B$_{28}$ allotrope.

\subsubsection*{Phosphorus}

\begin{figure}[]
\includegraphics[width=0.5\textwidth]{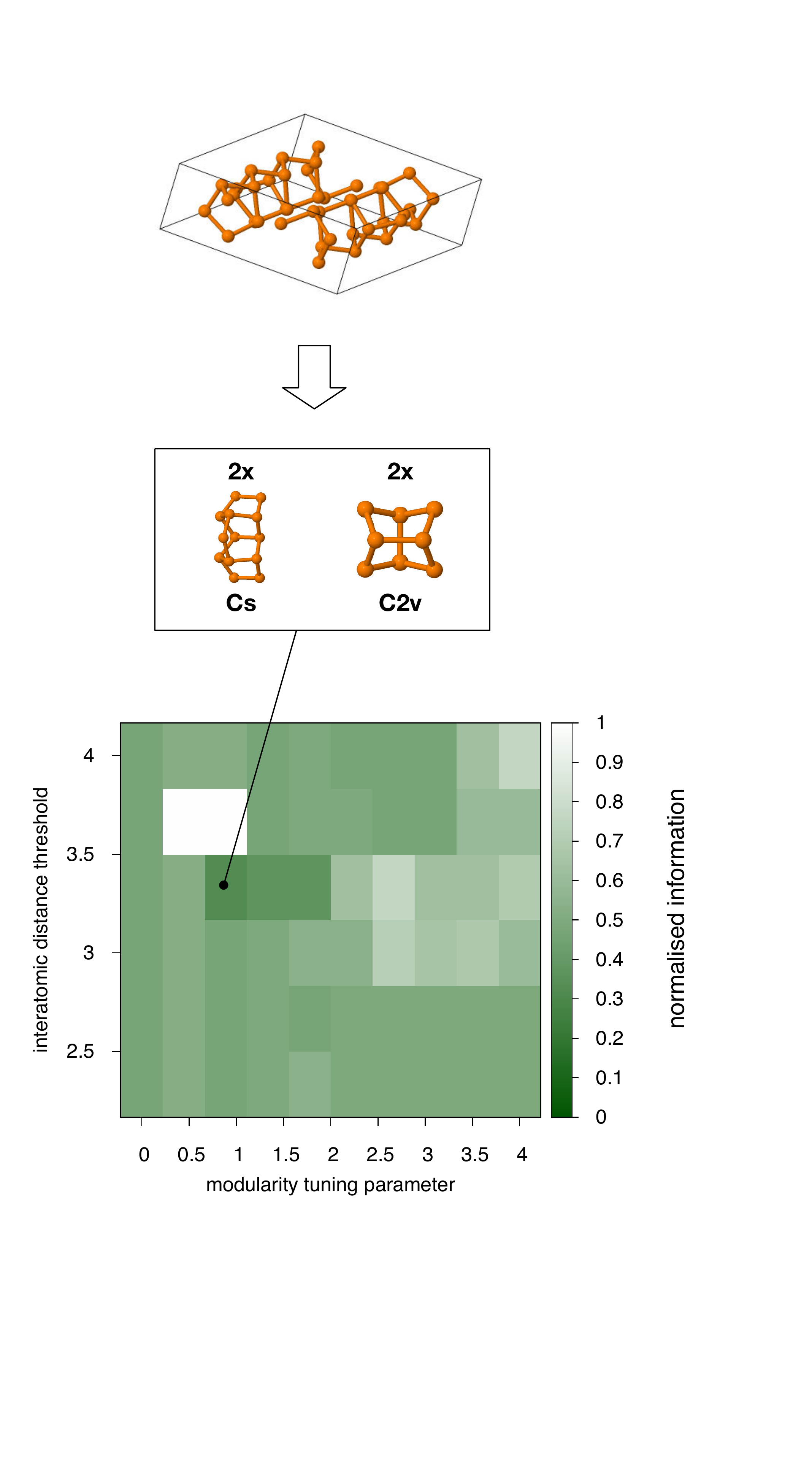}
\caption{The structure of red phosphorous (42 atom P$\bar 1$ unit cell) and its modular decomposition. The relatively low degree of symmetry in the structure as a whole results in a relatively flat landscape. The information $I$  (normalised here by the maximum value of $3N-6$) required to describe the structure is minimised for values around $w_s = 1$ and $d^* = 3.33$\AA. This decomposition consists of two modules, which each appear twice. One module consists of 13 atoms with C$_s$ point group symmetry. The other consists of eight atoms with C$_{2v}$ symmetry.}\label{P42}
\end{figure}

Like boron, phosphorous exhibits rich allotropism, from the highly metastable white phosphorous, to layered black phosphorous, and extremely complex fibrous, or layered, structures \cite{bachhuber2014extended}. There is considerable current interest in two dimensional black phosphorous, or phosphorene \cite{liu2014phosphorene} and other layered forms \cite{schusteritsch2016single}. Here we investigate the crystal structure of red phosphorus \cite{ruck2005faserformiger}, and attempt to identify a simple decomposition into modules using our current scheme. In the 42-atom primitive unit cell (space group P$\bar 1$) the modularity decomposition finds two modules that each occur twice (see Fig. \ref{P42}). The bigger module has symmetry (C$_s$) and contains 13 atoms, while the smaller has C$_{2v}$ symmetry and contains eight atoms. The relatively low degree of symmetry of red phosphorous means that the landscape of $I$ with varying $w_s$ and $d^*$ is flatter, but our approach nevertheless finds a parsimonious decomposition of the crystal structure.

\subsubsection*{Metal-organic frameworks}

We extend our framework to multi-species structures requiring only a definition of the relationship between $d^*$ and the interatomic distances used to determine the network connectivity. In principle, a separate $d^*$ could be defined for each pair of atomic species. However, this introduces the cost of exploring a higher dimensional space in the search for an optimal $I$. Instead, we define a single dimensionless parameter $d^*_{\rm eff}$ that specifies the distance threshold as the $d^*_{\rm eff}$-fold multiple of the sum of the fixed atomic radii for a given pair of atoms. We apply this multi-species version of our approach to the metal-organic framework MOF-5, or $\rm{Zn}_4\rm{O}(\rm{BDC})_3$, where $\rm{BDC}^{2-}$ is 1,4-benzenedicarboxylate \cite{li1999design}. Metal-organic frameworks exhibit a vast range of structures and are of great interest because their porosity allows them to be used for the storage of gases, such as hydrogen, or carbon dioxide \cite{kitagawa2014metal}. As can be seen in Fig. \ref{SAHYIK} our algorithm finds two similar decompositions with almost equally low $I$-values. The lowest minimum corresponds to a decomposition of the structure into six 16-atom modules with D$_{2h}$ symmetry, and two six-atom modules with tetrahedral symmetry (T$_d$). The 16-atom modules correspond to the $\rm{BDC}^{2-}$ molecules that are sometimes referred to as the 'struts' of metal-organic frameworks. This suggests that the decomposition derived automatically through our procedure is chemically meaningful.

\subsection*{Structure prediction}

Ab initio random structure searching (AIRSS) \cite{Pickard:2011bu,pickard2006high} is a simple, and demonstrably effective, approach to first principles structure prediction. It has been applied to a wide range of systems, from the crystal structures of  dense hydrogen \cite{pickard2007structure} and hydrogen rich compounds \cite{pickard2006high}, to matter under extreme compression \cite{pickard2010aluminium}, and interfaces \cite{schusteritsch2014predicting}. The approach involves selecting initially random structures from distributions defined by physically motivated constraints (for example, density, composition, atomic distances, symmetry, molecular units or fragments). These random ``sensible''\cite{Pickard:2011bu} structures are fully relaxed (moved to the nearest local minimum in the energy landscape) under forces derived from DFT. Once a large number of computations have been performed the resulting structures can be ranked according to energy (free enthalpy) or any computable property of interest.

It has been a surprise to many that such a naive approach performs well, but the method's success is linked to intrinsic features of the first principles energy landscape, such as its smoothness (a result of the quantum mechanical interactions between the atoms and electrons) and the relatively large number of large energy basins. In a smooth energy landscape the size of the basins correlates with their depth (deep basins occupying a large volume of configuration space), there is a natural bias in random sampling, plus relaxation, to the stable, low energy and relevant structures. In what follows we exploit our new approach for the automatic decomposition of known structures into minimum $I$ fragments to accelerate the search for complex structures by restricting the regions of configuration space that must be explored.

\subsubsection*{Application to dense boron}

\begin{figure}[]
\includegraphics[width=0.45\textwidth]{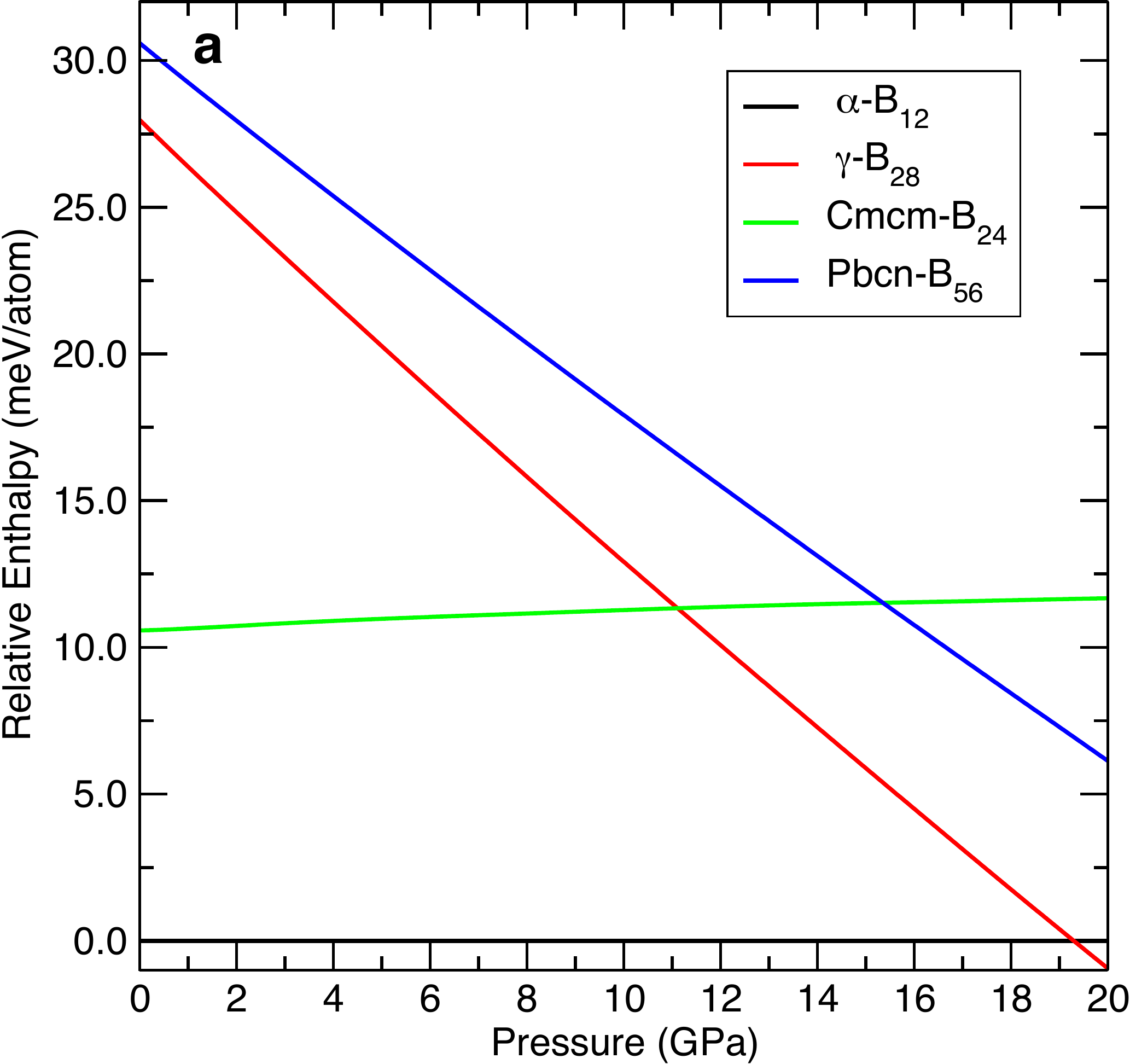}\\
\includegraphics[width=0.95\textwidth]{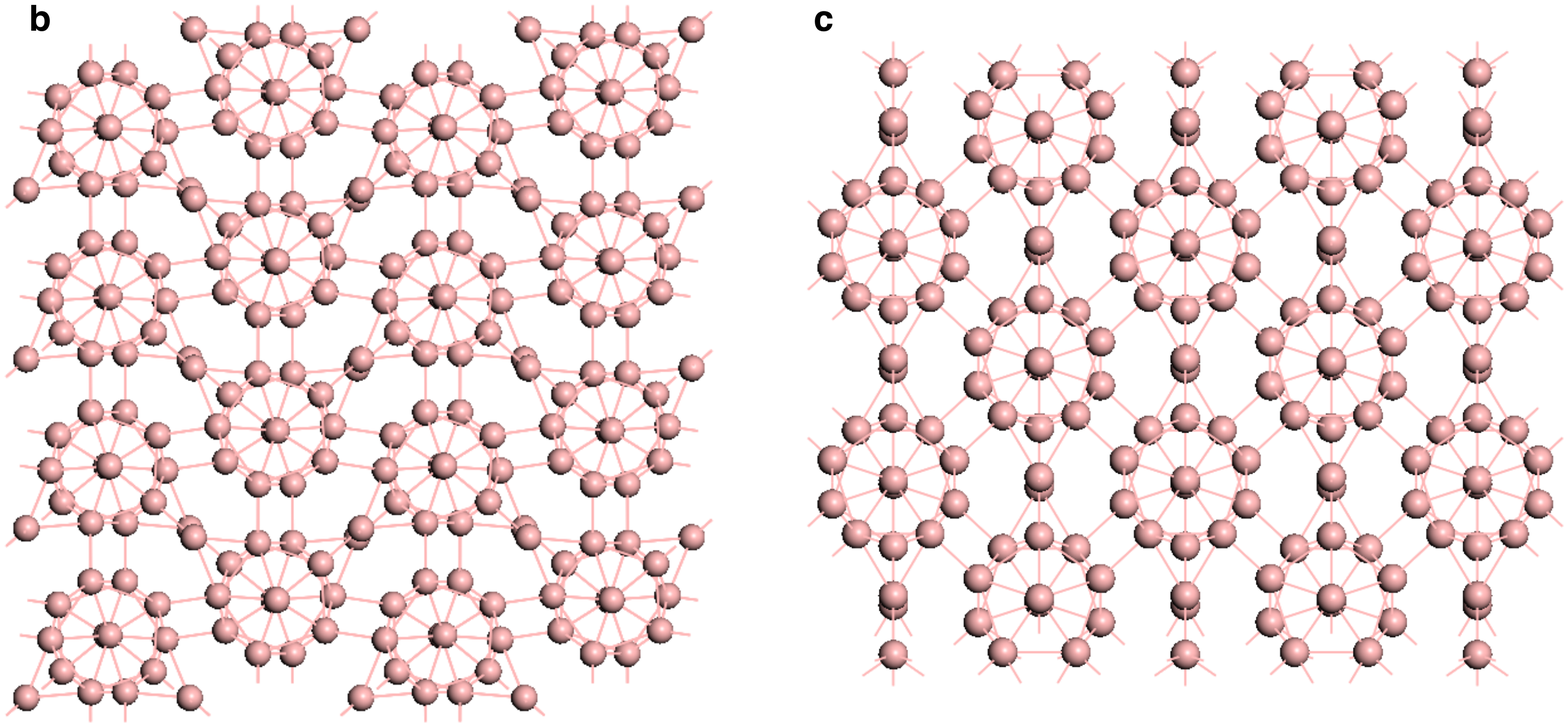}
\caption{(a) Relative enthalpy of boron structures up to 20GPa. (b) The new Pbcn-B$_{56}$ structure is practically degenerate in energy with (c) the known $\gamma$-B$_{28}$ structure.}
\label{newgp}
\end{figure}

\begin{figure}[]
\includegraphics[width=0.45\textwidth]{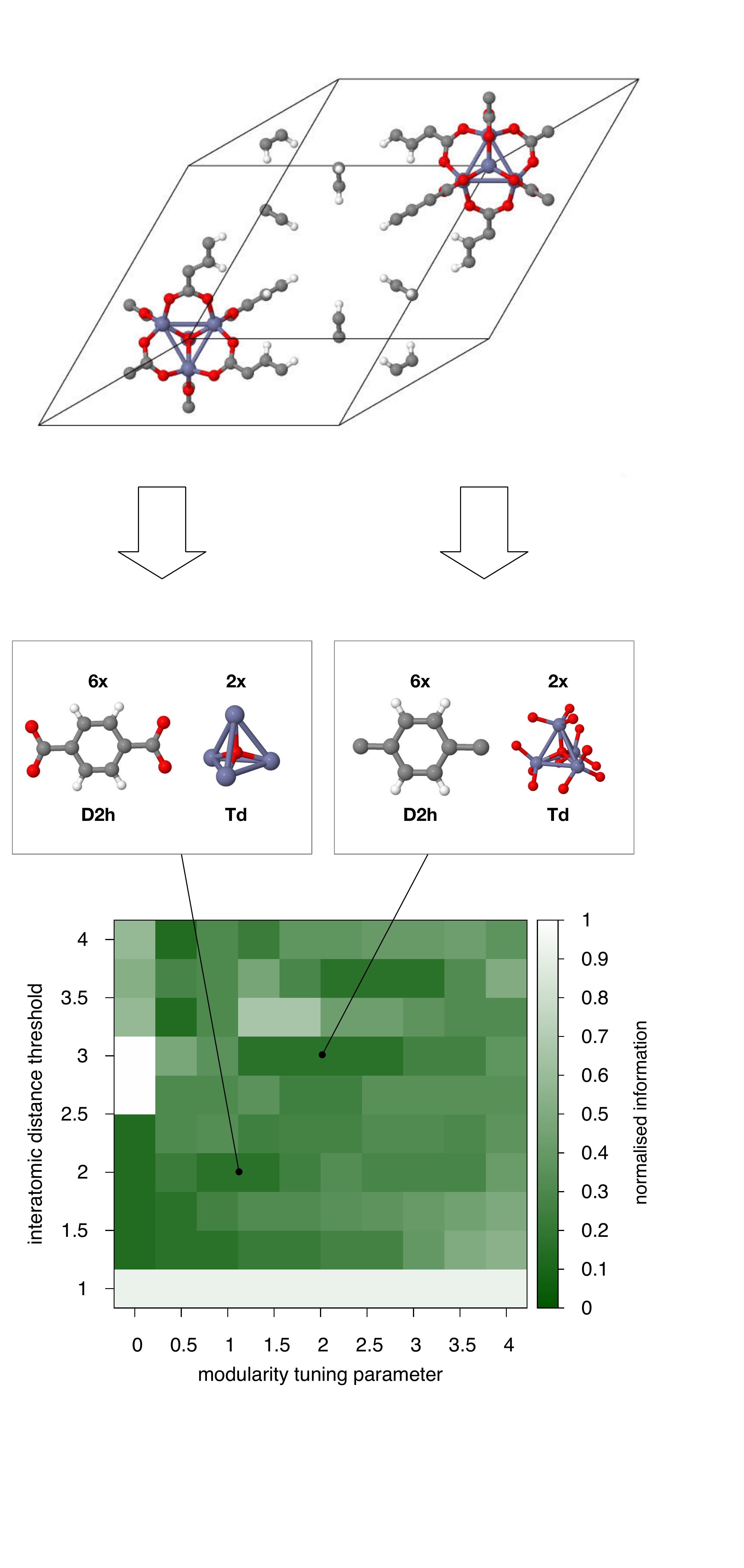}
\caption{The structure of the metal-organic framework MOF-5 ($\rm{Zn}_4\rm{O}(\rm{BDC})_3$, where $\rm{BDC}^{2-}$ is $1,4\rm{benzenedicarboxylate}$) and its modular decompositions. This structure has 106 atoms per unit cell. The two lowest minima have very similar values for the normalised information, with $I=0.163$ at $w_s = 2$ and $d^*_{\rm eff} = 3$ and $I=0.166$ at $w_s = 1$ and $d^*_{\rm eff} = 2$. Unlike $d^*$, which is a physical distance threshold, the distance threshold parameter $d^*_{\rm eff}$ is dimensionless and expressed as the multiple of the summed atomic radii for a given pair of atoms (see text). Both decompositions consist of two modules, one of which appears six times and has D$_{2h}$ symmetry, and the other one of which appears twice and has T${_d}$ symmetry. The modules for $I=0.163$ are one with 16 atoms (8 C, 4 O, 4 H) and one with five atoms (4 Zn, 1 O).}\label{SAHYIK}
\end{figure}

We generate 3303 initial random structures based on packing four 14 atom D$_{2h}$ modules derived from $\gamma$-B$_{28}$ into unit cells with a randomly chosen shape, and the same density as $\gamma$-B$_{28}$. The units are not permitted to overlap each other, or be closer than 1.63\AA (the measured inter-icosahedral distance in a computed $\alpha$-B structure), and are related to each other by symmetry. The symmetry is chosen at random from those space groups with four symmetry operators in the primitive cell. The random initial structures are then relaxed to nearby local energy minima (see Methods for computational details). Four of the initial structures relaxed to supercells of the Pnnm  $\gamma$-B$_{28}$ structure, and three of them relaxed to a previously unreported structure with space group Pbcn and 56 atoms in a unit cell. This structure has a density very close to that of $\gamma$-B$_{28}$, and is just 3 meV/atom less stable. In Fig. \ref{newgp} it is shown that this near degeneracy persists over a wide range of pressures. 

Our new 56 atom structure corresponds to a distorted hexagonal packing of boron icosahedra, whereas $\gamma$-B$_{28}$ corresponds closely to cubic packing. The small energy difference indicates that $\gamma$-B may be susceptible to polytypism or stacking disorder. The 3meV/atom energy difference between the hexagonal and cubic polytypes is small compared to the 25meV difference between cubic and hexagonal (carbon) diamond computed at the same level. The situation is very similar to that for $\alpha-$B$_{12}$, for which an alternatively packed structure of 24 atom and space group Cmcm has been identified \cite{Pickard:2011bu}, and further discussed \cite{he2013structures,zhu2015generalized}.

\section*{Discussion}

In contrast to other applications of community-detection algorithms, in which a degree of ambiguity in the definition of communities is often tolerated, our application relies on the detection of robust and compressible modules of atoms with a maximum amount of symmetry. Small differences in the assignments of atoms to modules can have large effects on the information measure $I$ if modular symmetries are established or broken as a result. For this reason the implementation of the modularity optimisation requires particular care in order to find maximally robust modular decompositions.

In the context of multi-species atomic structures, and particularly biological molecules, it can be valuable to consider weighted atomic networks, in which the edge weight scales with the interatomic distance. The simplest choice is a linear one:
\[
w_{ij} = K \left(1 - {d_{ij}\over d^*}\right)
\]
for $d_{ij} < d^*$ and $w_{ij} = 0$ otherwise,
where $d_{ij}$ is the distance between atoms $i$ and $j$, and $K$ is an arbitrary constant. Increasing all edge weights by a constant factor leaves the modularity unchanged, so in practice this is chosen to ensure numerical precision. This choice of edge weighting reflects the fact that equilibrium bond lengths scale with the equilibrium bond energies.

Our approach can be applied in the context of proteins, similar to \cite{Delvenne:2010iw,Delmotte:2011bn,Yaliraki:2014wj}, where the vastness of configuration space has traditionally also been a difficult barrier to overcome. The lack of symmetry in biological molecules mean that the information required to describe the structure is less useful than in crystal structures. Other measurements, such as the stability of the module number, can replace the information measurement as a criterion for assessing the quality of a modular decomposition in biomolecules. While methods for rigidity analysis in proteins already exist, such as the FIRST algorithm \cite{Jacobs:2001we}, the tuning parameter in our method allows for the detection of rigid clusters on a variety of length scales, making it a complementary approach. Like FIRST, our method could inform coarse-grained multi-scale molecular dynamics methods such as FRODA \cite{Wells:2005pb}, supplying the rigid units described as 'ghost templates' in FRODA, and lead to more efficient computational models of protein dynamics.

In the context of complex crystal structures this approach has numerous potential applications. It first of all suggests an automatic coarse-graining and thereby provides an intuitive simplification and visual aid. The discovery of modules also facilitates the accelerated exploration of configuration space, particularly in the context of random structure search. This has been demonstrated by the new structure of boron closely related to $\gamma$-B$_{28}$ that we find using a module-based search. We believe that these results show the potential of atomic network analysis as a tool for materials discovery. Our algorithm is evidently fast enough for these purposes - the run time for calculating the metal-organic framework MOF-5 decomposition (106 atoms per unit cell) was 2.3 seconds on a 3.1 GHz Intel Core i7 processor.

\section*{Methods}

The density functional computations on boron were performed using CASTEP 17.2 \cite{clark2005first}, a full featured plane wave pseudopotential total energy code. The GGA-PBE \cite{perdew1996generalized} density functional was used, along with a legacy Vanderbilt ultrasoft pseudopotential \cite{vanderbilt1990soft}, a plane wave cutoff of 240 eV, and a k-point sampling density of 0.1$\times2\pi$\AA$^{-1}$, for the random searches. The enthalpies were calculated using a higher precision default on-the-fly pseudopotential, 700 eV plane wave cutoff and a k-point sampling density of 0.03$\times2\pi$\AA$^{-1}$. 

\section*{Author Contributions }

SEA and CJP designed the study and methodology. SEA, CJP and WPG performed computations, generated data, and wrote the manuscript.

\section*{Author Information}

The authors declare no competing financial interests. 

\section*{Data Availability}

The data associated with this manuscript is made available in Ref. \cite{Pickard2017}

\section*{Acknowledgments}

SEA is supported by a Royal Society University Research Fellowship and a Gatsby Career Development Fellowship. CJP acknowledges financial support from the Engineering and Physical Sciences Research Council (EPSRC) of the United Kingdom (Grant Nos. EP/G007489/2 and EP/K013688/1). CJP is also supported by the Royal Society through a Royal Society Wolfson Research Merit award. WPG acknowledges financial support from the EPSRC Centre for Doctoral Training in Computational Methods for Materials Science under grant EP/L015552/1.


\hfill
\newpage
\hfill
\newpage

\end{document}